\documentclass[useAMS,usenatbib]{mn2e}
\usepackage{graphicx}
\usepackage{times}
\usepackage{epsfig}
\usepackage{longtable}
\usepackage{lscape}
\usepackage{amssymb}
\usepackage{latexsym}
\usepackage{natbib}
\def\kms{$\rm km\;s^{-1}$}

\def\hb{H$\beta$}
\def\hi{H{\sc~i}}

\def\mgb{Mg{\it b}}

\def\fei{Fe{\small 5270}}
\def\feii{Fe{\small 5335}}
\def\fe{$\langle\,{\rm Fe}\rangle$}

\title[Counter-rotating discs in NGC 5719]{Dating the formation of the
  counter-rotating stellar disc in the spiral galaxy NGC 5719 by
  disentangling its stellar populations\thanks{Based on observations
    collected at the European Southern Observatory for the program
    383.B-0632.}}

\author[L. Coccato et al.]{L. Coccato$^{1}$\thanks{E-mail:
    lcoccato@eso.org}, L. Morelli$^2$, E. M. Corsini$^2$,
  L. Buson$^3$, A. Pizzella$^2$, D. Vergani$^4$, \and F. Bertola$^{2}$
\\$^{1}$European Southern Observatory, Karl-Schwarzschild-Stra$\beta$e
2, D-85748 Garching bei M\"unchen, Germany.  
\\$^{2}$Dipartimento di Astronomia, Universit\`a di Padova, vicolo
dell'Osservatorio 3, I-35122 Padova, Italy .
\\$^3$INAF, Osservatorio Astronomico di Padova, vicolo
dell'Osservatorio 5, I-35122 Padova, Italy.
\\$^4$INAF, Osservatorio Astronomico di Bologna, via Ranzani 1,
I-40127, Bologna, Italy.}

\begin{document}

\date{Accepted... Received...}

\pagerange{L\pageref{firstpage}--L\pageref{lastpage}} \pubyear{2011}

\maketitle

\label{firstpage}

\begin{abstract}
We present the results of the VLT/VIMOS integral-field spectroscopic
observations of the inner $28''\times28''$ ($3.1\;{\rm kpc} \times
3.1\;{\rm kpc}$) of the interacting spiral NGC~5719, which is known to
host two co-spatial counter-rotating stellar discs.
At each position in the field of view, the observed galaxy spectrum is
decomposed into the contributions of the spectra of two stellar and
one ionised-gas components. We measure the kinematics and the line
strengths of the Lick indices of the two stellar counter-rotating
components.  We model the data of each stellar component with single
stellar population models that account for the $\alpha$/Fe
overabundance.
We also derive the distribution and kinematics of the ionised-gas
disc, that is associated with the younger, less rich in metals, more
$\alpha$-enhanced, and less luminous stellar component. They are both
counter-rotating with respect the main stellar body of the galaxy.
These findings prove the scenario where gas was accreted first by
NGC~5719 onto a retrograde orbit from the large reservoir available in
its neighbourhoods as the result of the interaction with its companion
NGC~5713, and subsequently fuelled the {\em in situ\/} formation of
the counter-rotating stellar disc.

\end{abstract}


\section{Introduction}

The presence of stars counter-rotating with respect to other stars
and/or gas has been detected in several disc galaxies and is commonly
interpreted as the end result of a retrograde acquisition of external
gas and subsequent star formation (see \citealt{Bertola+99b} for a
review). Nevertheless, some special cases of counter-rotating stellar
discs could have an internal origin induced by the presence of a bar
(e.g., \citealt{Evans+94}).

The demography of gaseous and stellar counter-rotating components in
S0's and spirals is a key to understand their assembly process.  The
fraction of lenticular galaxies with a counter-rotating gaseous disc
is consistent with the 50\% that we expect if all the gas in S0's is
of external origin \citep{Bertola+92}. In contrast, less than 10\% of
them host a detectable fraction of counter-rotating stars
\citep{Kuijken+96}.  Large-scale counter-rotation is a rare phenomenon
in spirals. In fact, less than 10\% of the studied spiral galaxies
host a counter-rotating gaseous and/or stellar disc
\citep{Kannappan+01, Pizzella+04}. The particular case of large-scale
stellar counter-rotation in disc galaxies has been observed only in
NGC 4550 \citep{Rubin+92, Rix+92, sauron3}, NGC~7217
\citep{Merrifield+94}, NGC~3593 \citep{Bertola+96, Corsini+98b,
  Garcia+00}, NGC~4138 \citep{Jore+96}, and NGC~5719
\citep{Vergani+07}.
To interpret the observed frequencies of counter-rotations,
\citet{Pizzella+04} argue that the retrograde acquisition of small
amounts of external gas can give rise to counter-rotating gaseous
discs in gas-poor S0's only, while in gas-rich spirals the newly
acquired gas is swept away by the pre-existing gas. 
  Counter-rotating gaseous discs in spirals are formed only from the
  retrograde acquisition of amounts of gas larger than the
  pre-existing gas content. This is the case of the purely gaseous
  counter-rotating components detected in NGC~3626 \citep{Ciri+95,
    Garcia+98} and NGC~4826 \citep{Braun+94, Garcia+03}.
  Counter-rotating stellar discs are produced by subsequent star
  formation. Therefore, in this scenario counter-rotating stellar
  discs are expected to be made by younger stars than those of their
  host galaxy.

This picture can be directly tested in the NGC~5719/13 galaxy pair,
which has been recently studied by \citet{Vergani+07}. NGC~5719 is an
almost edge-on Sab galaxy with a prominent skewed dust lane at a
distance of 23.2 Mpc. \citet{Vergani+07} report a spectacular on-going
interaction with its face-on Sbc companion NGC~5713. Two \hi\ tidal
bridges loop around NGC~5719 and connect it to NGC~5713 at a projected
distance of 77 kpc.
The neutral and ionised hydrogen in the disc of NGC~5719 are
counter-rotating with respect to the main stellar disc. 
%
%
The kinematics of the ionized-gas disc and of both the
  counter-rotating stellar discs are measured out to about $40''$ (4.3
  kpc) from the galaxy centre.
In conclusion, \citet{Vergani+07} propose a scenario where \hi\ from
the large reservoir available in the galactic surroundings was
accreted by NGC~5719 onto a retrograde orbit and subsequently fuelled
the {\em in situ\/} formation of the counter-rotating stellar disc.

In this work, we will address the crucial piece of information which
is still missing, i.e. proving that the stellar population of the
counter-rotating disc of NGC~5719 is younger with respect to that of
the stars in the galaxy main disc.

\begin{figure}
\hspace{-0.5cm}
\psfig{file=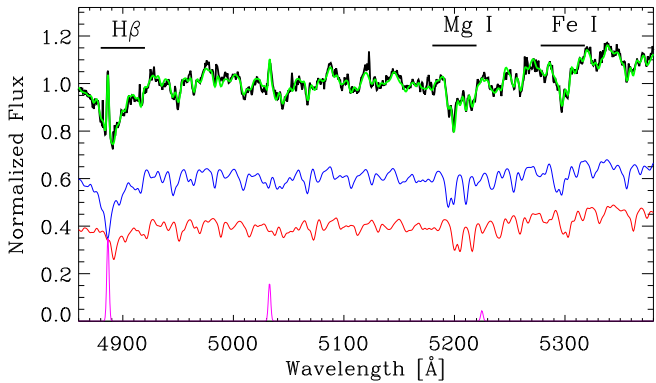,width=8.7cm,clip=}
\caption{Fit of the galaxy spectrum ({\it black}) in the spatial bin
  at $10''$ East from the galaxy center. The best fit model ({\it green}) is the
  sum of the spectra of the ionised-gas component ({\it magenta}) and the
  two stellar components ({\it blue} and {\it red}). The latter are obtained
  convolving the synthetic templates with the best fitting Gaussian
  LOSVDs and multiplying them by the best fitting Legendre
  polynomial. The differences in the position of absorption line
  features and in the \hb\ equivalent widths between the two stellar
  components (indicating different kinematics and stellar population
  content) are clearly evident.}
\label{fig:kinem_fit}
\end{figure}

\section{Observations and Data reduction}
\label{sec:1}

The integral-field spectroscopic observations were carried out in
service mode with the Very Large Telescope (VLT) at the European
Southern Observatory (ESO) in Paranal during dark time between 28 April
and 16 June 2009. The Unit Telescope 3 was equipped with the
Visible Multi Object Spectrograph (VIMOS) in the Integral Field Unit
(IFU) configuration. The HR blue grism covering the spectral range
4150 -- 6200 \AA\ and the 0\farcs67 fibre$^{-1}$ resolution were used.
The instrumental spectral resolution measured at 5200 \AA\ was $2.0$
\AA\ (FWHM), equivalent to 115 \kms.
Observations were organised into three dithered on-target exposures of
2950 seconds each, alternated to three offset sky exposures of 280
seconds each. The average seeing measured by the ESO Differential Image
Meteo Monitor was $1''$.

Data reduction (bias subtraction, fibre identification and tracing,
flat fielding, wavelength calibration and correction for instrument
transmission) was performed using the VIMOS ESO pipeline version 2.2.1
 (http://www.eso.org/sci/software/pipelines/).  The different
relative transmission of the VIMOS quadrants was corrected by
comparing the intensity of the night-sky emission lines. The three
offset observations were used to construct three sky spectra. To
compensate for the time variation of the relative intensity of the
night-sky emission lines, we compared the fluxes of the sky lines
measured in the offset and on-target exposures. The corrected sky
spectra were then subtracted from the corresponding on-target
exposures. Each exposure was organised in a data cube using the
tabulated correspondence between each fibre and its position in the
field of view. The three sky-subtracted data cubes were aligned using
the bright galaxy nucleus as reference and co-added into a single data
cube.

In order to increase the signal to noise ratio ($S/N$), spectra from
fibres mapping adjacent regions in the sky were added together using
the Voronoi binning method \citep{Cappellari+03}.  Some of the spatial
bins were modified to include only spectra from the regions associated
to intense \hb\ emission or the dust lanes crossing the galactic disc.
This ensured us to minimise the contamination of the kinematic and
stellar-population properties to be measured from spectra obtained in
regions where stellar counter-rotation was detected and chemical
decoupling is expected \citep{Vergani+07}. In total, we have 130
spatial bins few arcseconds square large. We tested the robustness of
our results using different binning schemes.

\begin{figure*}
\hspace{-0.7cm}
\hbox{
\psfig{file=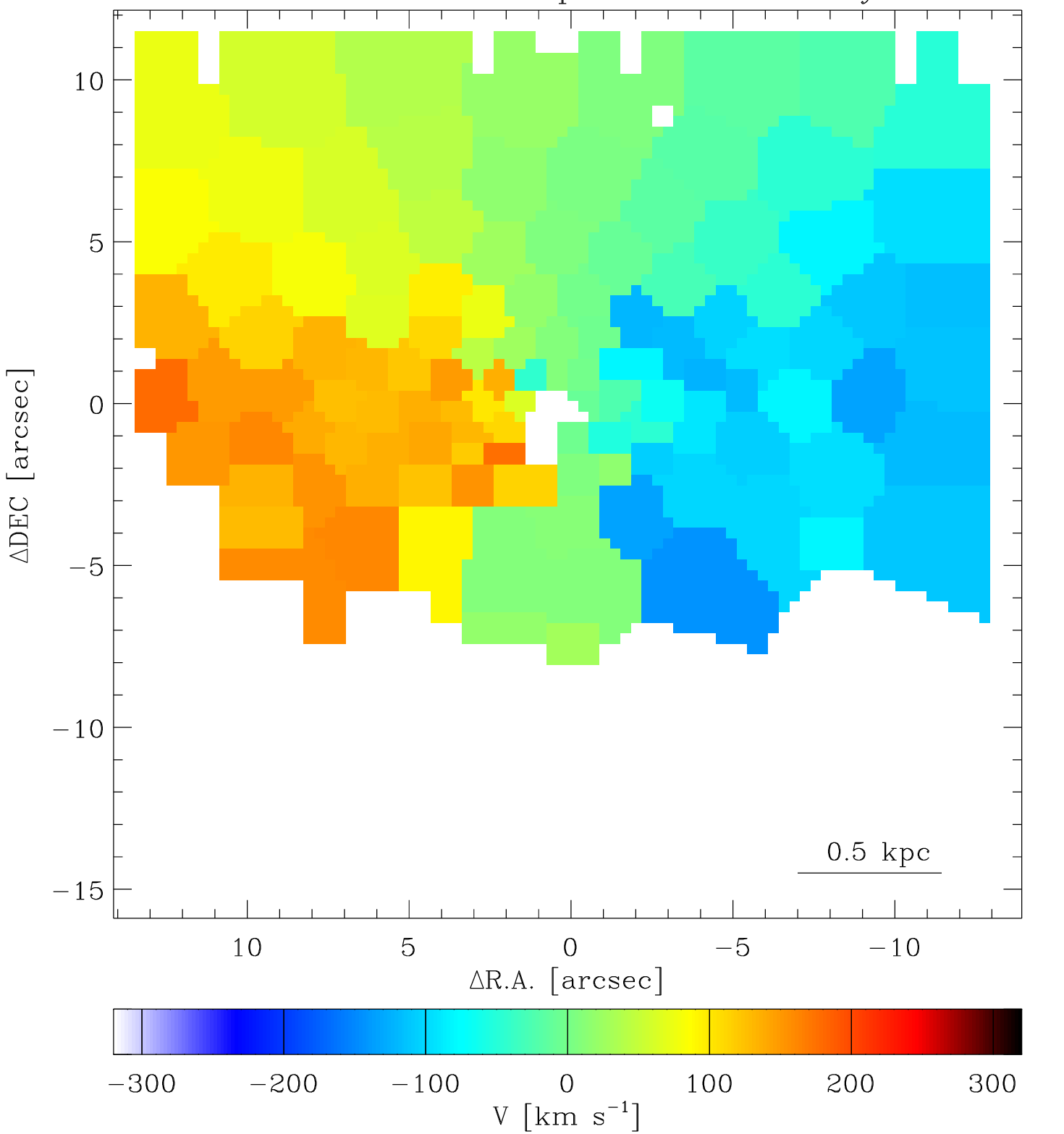,clip=,width=5.9cm,bb=53 359 467 828}
\psfig{file=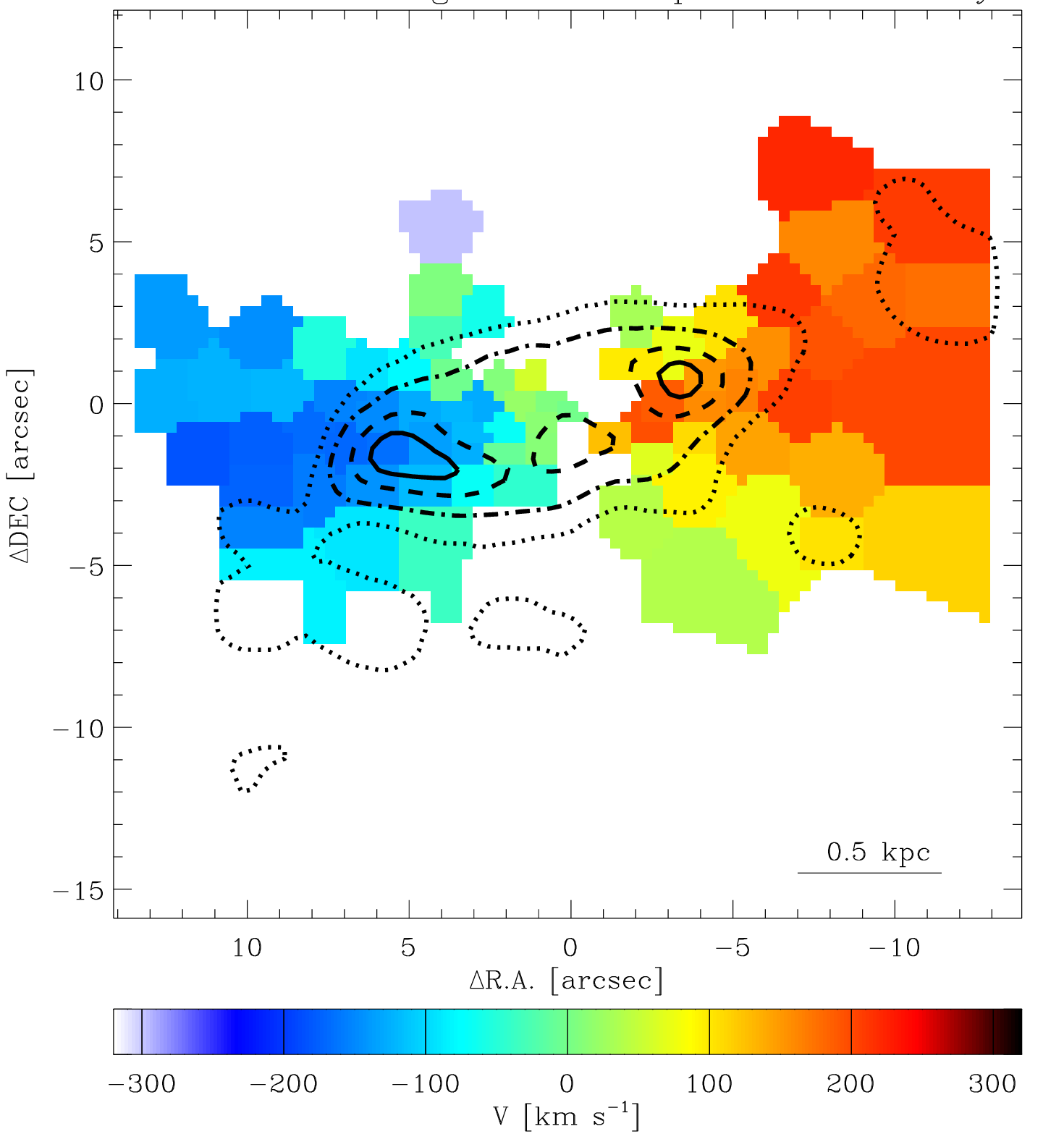,clip=,width=5.9cm,bb=53 359 467 828}
\psfig{file=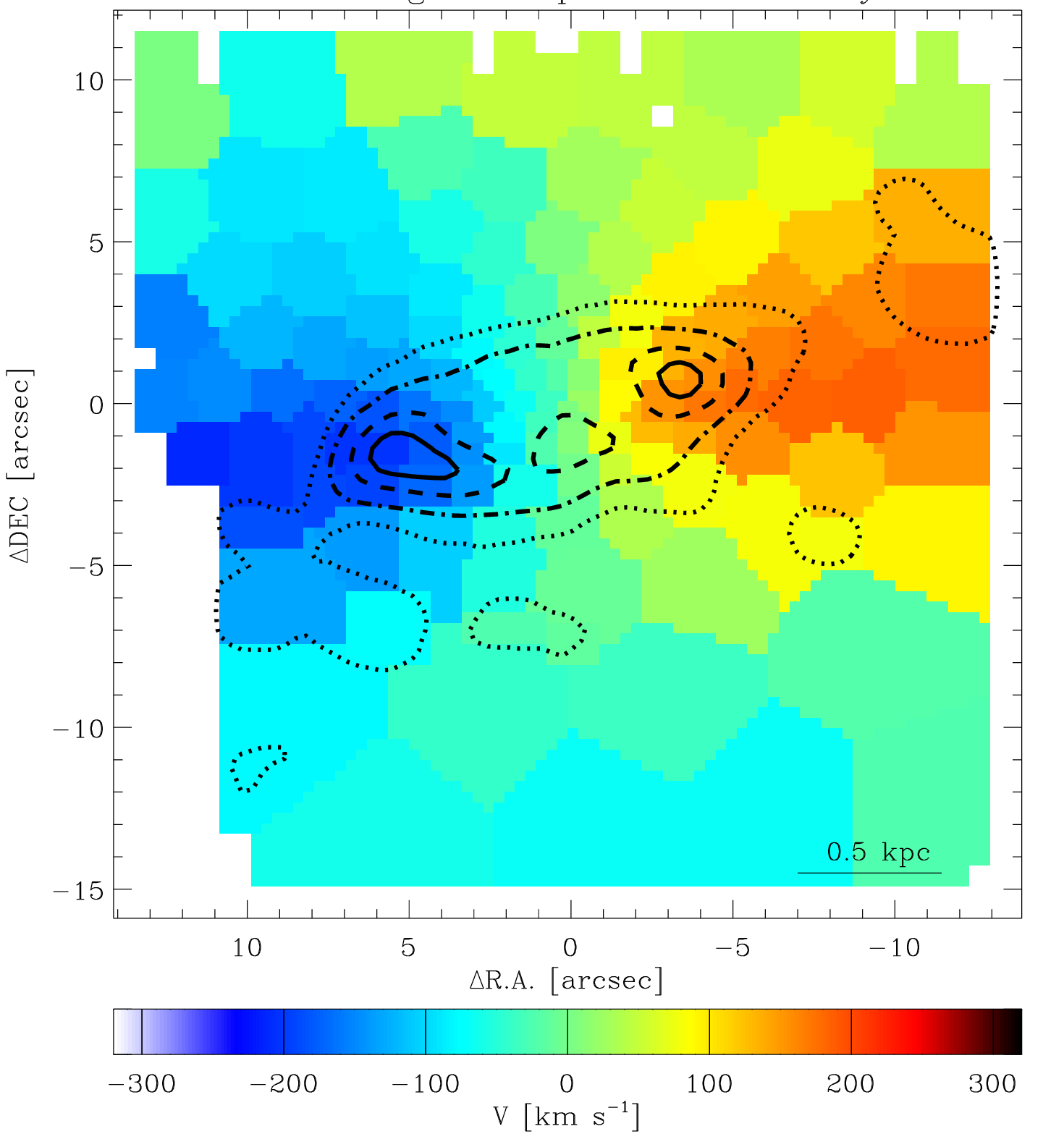,clip=,width=5.9cm,bb=53 359 467 828}
}
\caption{Velocity field of the main stellar component ({\it left panel}),
  counter-rotating stellar component ({\it central panel}) and
  counter-rotating ionised gas ({\it right panel}) in NGC 5719. Centre is
    in R.A. = 14$^{\rm h}$40$^{\rm m}$56.3$^{\rm s}$,
    Dec. = $-$00$^\circ$19$'$05.4$''$ (J2000.0). {\it Black contours:} isophotes derived from the \hb\ emission map. }
\label{fig:2dkin}
\end{figure*}

\begin{figure*}
\hspace{-0.7cm}
\psfig{file=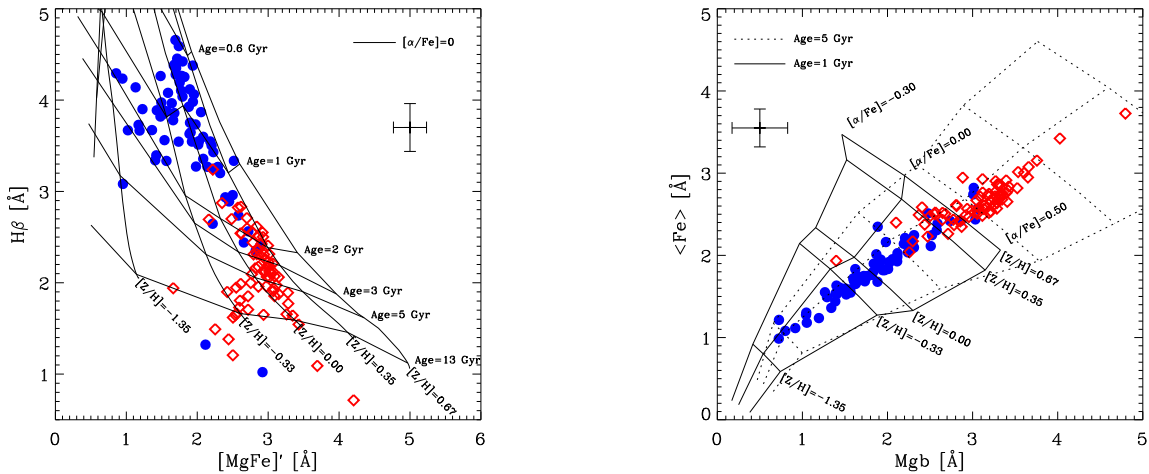,width=15.7cm,clip=}
\caption{Equivalent width of the Lick indices in the main ({\it red
    diamonds}) and counter-rotating ({\it blue
    circles}) stellar components. Predictions from single stellar population models by
  \citet{Thomas+10} are superimposed. {\it
    Crosses} indicate mean error bars associated to the equivalent widths.}
\label{fig:indices}
\end{figure*}

\section{Kinematics and line strength indices}
\label{sec:2}

In order to measure the kinematics and stellar population properties
of the two counter-rotating stellar components in NGC 5719, we need to
separate their contribution to the observed spectrum in each spatial
bin, taking advantage of their different velocities that causes a
wavelength shift of their spectra. To do that, we modified the 
  penalized pixel fitting code (pPXF, \citealt{Cappellari+04}) in a
similar way to that done by \citet{McDermid+06}. The code builds two
synthetic templates (one for each stellar component) as linear
combination of stellar spectra from the MILES library (at ${\rm
  FWHM}=2.54$ \AA\ spectral resolution, \citealt{Beifiori+10}) and
convolves them with two Gaussian line-of-sight velocity distributions
(LOSVDs) with different radial velocities and the same velocity
dispersion $\sigma$.
This assumption is in agreement with the findings by
\citet{Vergani+07}. Results are confirmed if two different velocity
dispersions are considered, although the velocity fields and maps of
stellar population properties appear more noisy due to the
additional degree of freedom that increases the degeneracy between
measured parameters.
Gaussian functions are added to the convolved synthetic templates to
account for ionised-gas emission lines (H$\beta$, [O{\sc\
    iii}]$\lambda\lambda$4959, 5007, and [N{\sc\ i}]$\lambda$5198) and
fit simultaneously to the observed galaxy spectra.
Multiplicative Legendre polynomials are included to match the shape of
the galaxy continuum, and are set to be the same for the two
synthetic templates.
Our technique represents an improvement with respect to
\citet{Vergani+07}: in fitting the measured LOSVD with a double
Gaussian, they assumed that the two counter-rotating components had the
same stellar population. On the contrary, our synthetic templates
account for different stellar populations since they are
independently built from the MILES library.
We check the robustness of our results using also the INDO-US
  Coud\'e library of stellar spectra \citep{Valdes+04} and the single
  stellar population (SSP) synthesis models by \citet{Vazdekis+10} to
  build the synthetic templates.
Figure \ref{fig:kinem_fit} shows an example of the decomposition of
the galaxy spectrum measured in one of the spatial bins where the two
stellar counter-rotating components and ionised gas are observed.
We perform the double stellar component fit only in 73 spatial bins,
where the separation of the two kinematic components is reliable, and
do the single stellar component fit to the spectra of the other
bins. The spectra of outermost spatial bins have a low $S/N$,
therefore only the ionised-gas emission lines are fitted.  The stellar
kinematics are measured by taking into account the difference in
spectral resolution between the VIMOS spectra and the MILES library.
We derive the properties of the stellar populations of the two
  stellar counter-rotating components of NGC 5719 by measuring the
  line strength of the Lick indices \hb , \mgb , \fei , and \feii\ as
  defined by \citet{Worthey+94} in the synthetic templates.
To this aim, the spectra are set to rest-frame by adopting the
measured radial velocity and convolved with a Gaussian function to
match the spectral resolution of the Lick system ($\rm FWHM = 8.4$
\AA).
We also calculate the mean iron index \fe\ = $\left( {\rm
    \fei} + {\rm \feii} \right)/2$ \citep{Gorgas+90} and the combined
  magnesium-iron index [MgFe]$' = \sqrt{{\rm \mgb} \left( 0.82 \cdot
    {\rm \fei} + 0.28 \cdot {\rm \feii} \right)}$. The [MgFe]$'$ index
  is almost independent from $\alpha$-enhancement and hence serves
  best as a metallicity tracer \citep{Thomas+03}.

We assume that the two synthetic templates are a good approximation of
the mean stellar populations of the two counter-rotating
components. This depends on the adopted library of stellar spectra and
on the accuracy of the fitting program in recovering {\em
  simultaneously\/} the kinematics (i.e, radial velocity and velocity
dispersion) {\em and\/} the population properties (i.e., the continuum
and lines strength) of the two counter-rotating components.
\begin{figure*}
\hspace{-0.5cm}
\vbox{
  \hbox{
    \psfig{file=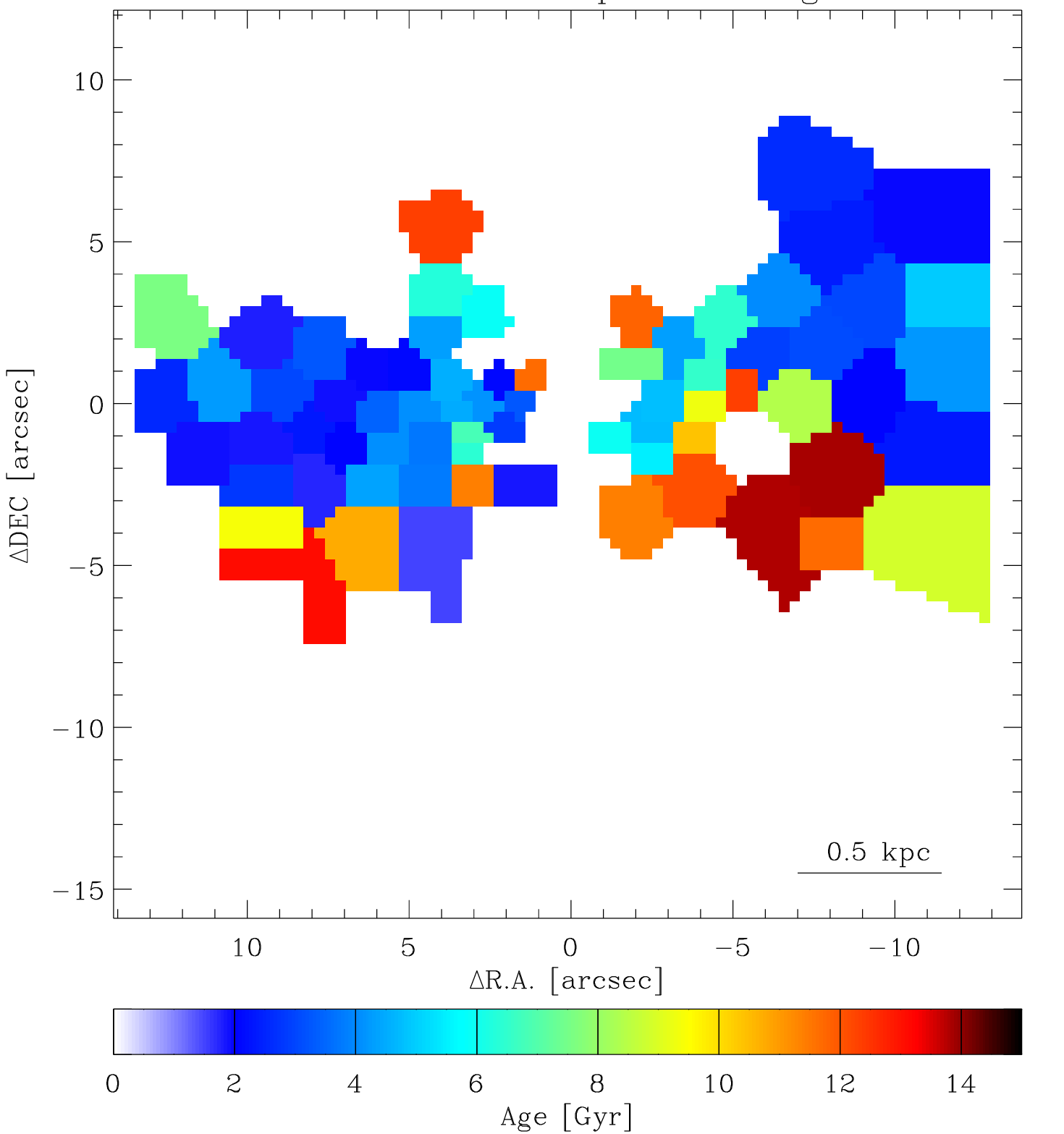,clip=,width=5.9cm,bb=53 359 467 828}
    \psfig{file=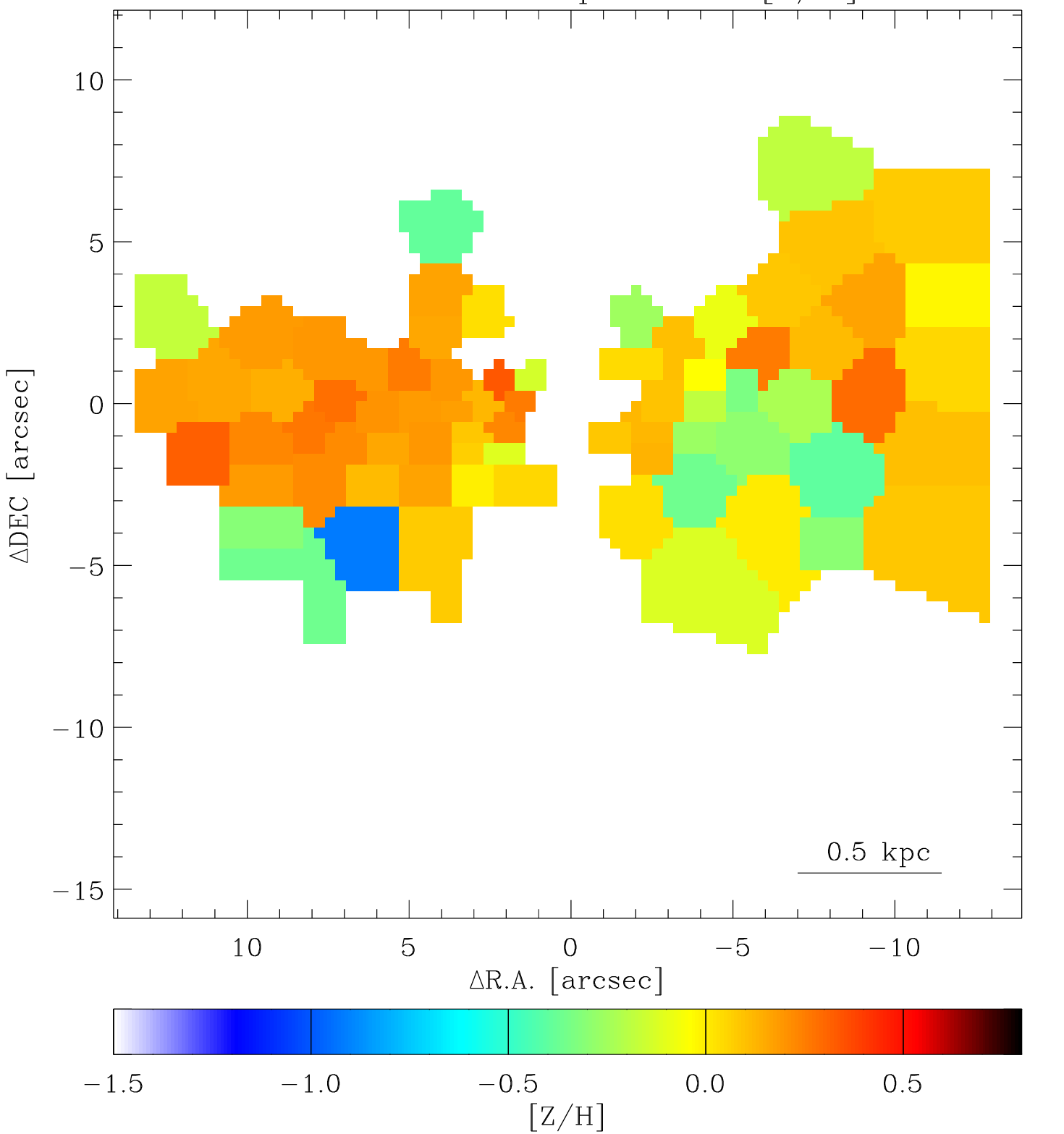,clip=,width=5.9cm,bb=53 359 467 828}
    \psfig{file=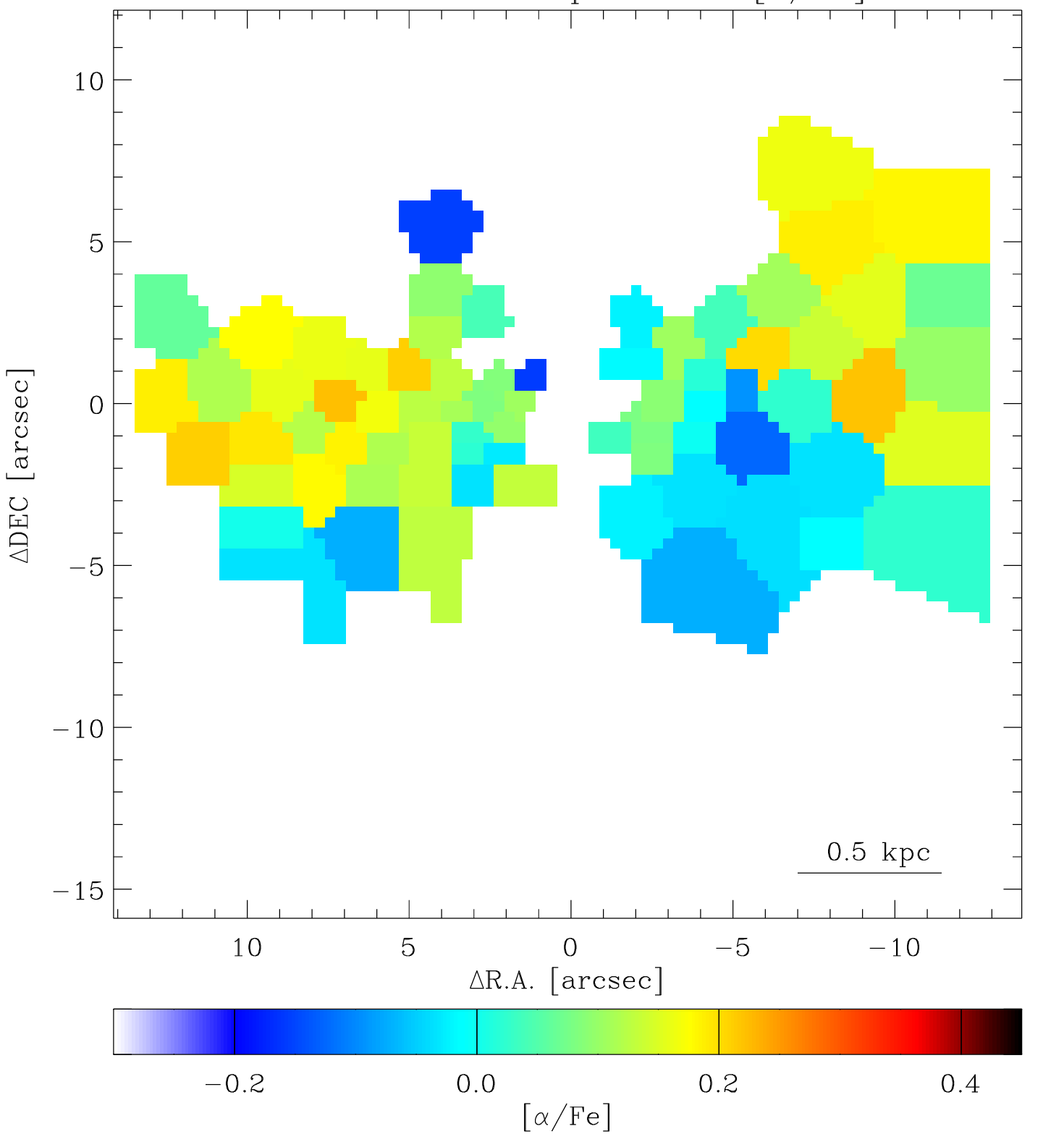,clip=,width=5.9cm,bb=53 359 467 828}
  }
\vspace{0.1cm}
  \hbox{
    \psfig{file=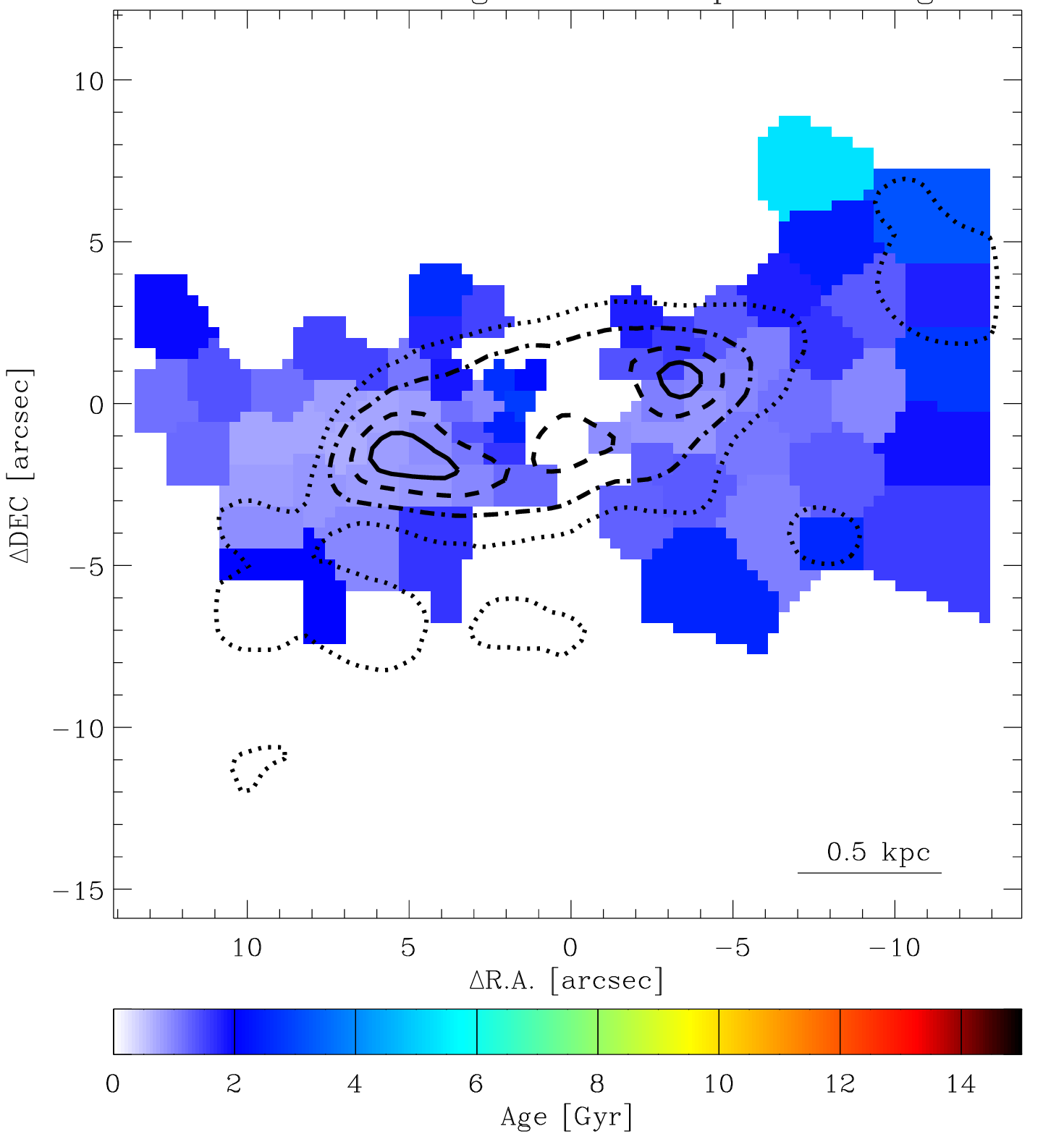,clip=,width=5.9cm,bb=53 359 467 828}
    \psfig{file=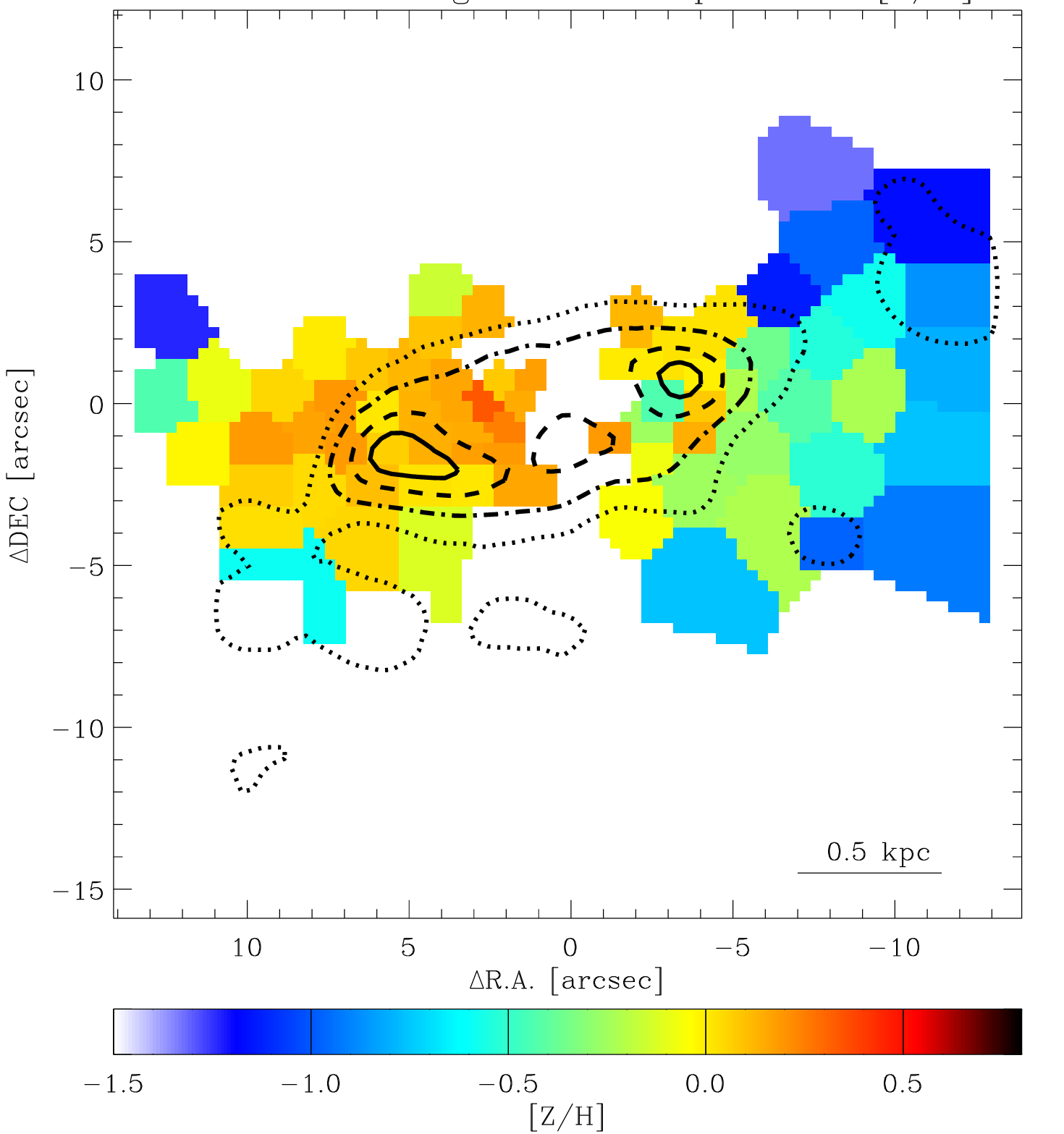,clip=,width=5.9cm,bb=53 359 467 828}
    \psfig{file=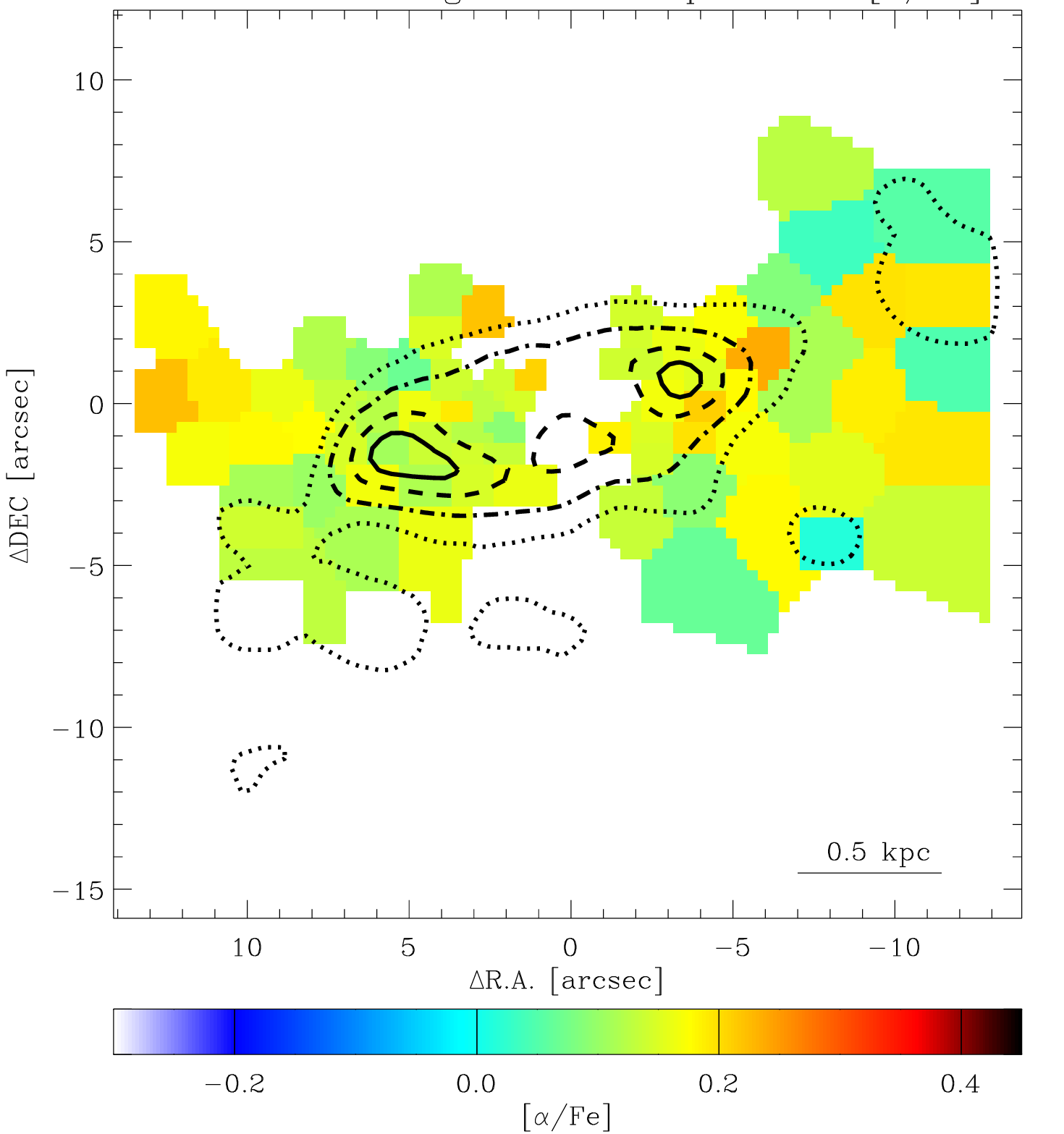,clip=,width=5.9cm,bb=53 359 467 828}
  }
}
\caption{Two-dimensional maps of age, [Z/H] and [$\alpha$/Fe] of the
  main ({\it upper panels}) and counter-rotating ({\it lower panels}) stellar
  components, in the region where the two components are
  disentangled. {\it Black contours:} isophotes derived from
  the \hb\ emission map. Centre is as in Fig. \ref{fig:2dkin}.}
\label{fig:2dssp}
\end{figure*}
To test this, we carry out Monte Carlo simulations on a set of 1200
artificial galaxy spectra to test the reliability and accuracy of the
procedure to measure the kinematics and line strength of the Lick
indices of the two counter-rotating components and quantify the
errors.  We generate each galaxy spectrum by summing two stellar
spectra with different velocity separation ($\Delta V$), different
fraction of stars in the main component with respect to the total
($F_{\rm main}$), different $\sigma$, and different $S/N$.  The two
stellar spectra are chosen to have similar line strengths to those
measured for the corresponding stellar component. Gas emission lines
are also added to the artificial galaxy spectra.
The modified pPXF routine is then applied to
analyse the artificial galaxy spectra as if they are real.
Errors on the fitted parameters are estimated by comparing the input
and output values. Errors are assumed to be normally distributed, with
mean and standard deviation corresponding to the systematic
($\Delta_{\rm S}$) and random ($\Delta_{\rm R}$) errors on the
relevant parameters, respectively.
We find that for typical values observed in NGC~5719 ($\Delta V
\gtrsim 150$ \kms, $30 \lesssim S/N \lesssim 90$, $\sigma \simeq 90$
\kms, $30\% \lesssim F_{\rm main} \lesssim 70\%$) the random errors on
the recovered $\Delta V$ and $F_{\rm main}$ are smaller than 25
\kms\ and 11\%, respectively, while systematic errors are
negligible. The measurements of the line strengths are affected both
by random and systematic errors: if the true equivalent width of the
spectral index $I$ of the two components are $I_1$ and $I_2$ with $I_1
< I_2$, the program measures $\left( I_1+\Delta_{\rm S} \right) \pm
\Delta_{\rm R}$ and $\left( I_2-\Delta_{\rm S} \right) \pm \Delta_{\rm
  R}$. For the typical values observed in NGC 5719, random errors are
$0.15-0.30$ \AA, and systemic errors are $0.10-0.20$ \AA, mildly
depending on $I_1$ and $I_2$.
Figure \ref{fig:2dkin} shows the measured two-dimensional velocity
fields of the main and counter-rotating stellar components and ionised
gas. Figure \ref{fig:indices} shows the Lick indices [MgFe]$'$, \hb,
\mgb, and \fe\ we measure in the bins where the two counter-rotating
components are disentangled. The predictions of the single stellar
population models by \citet{Thomas+10} are also shown for reference.
Correction for the offset to the Lick system is not applied,
  since no standard Lick stars were observed. Nevertheless, the VIMOS
  spectra are flux calibrated, therefore we expect the offset to be
  small and constant. This represents only a rigid shift to the points
  in Fig. \ref{fig:indices}, and therefore does not change our
 result.

The luminosity-weighted metallicity ([Z/H]), $\alpha$-enhancement
([$\alpha$/Fe]) and population age are derived for each component by a
linear interpolation between the model predictions by
\citet{Thomas+10} using an iterative procedure as done in
\citet{Morelli+08}.  We do not include in the interpolation all the
spatial bins with equivalent width of the Lick indices located outside
the model grid more than the mean errorbar.  The two-dimensional maps
of age, [Z/H] and [$\alpha$/Fe] of the main and counter-rotating
stellar components are shown in Figure \ref{fig:2dssp}.

\section{Discussion and conclusions}
\label{sec:3}

We present the results of the VLT/VIMOS integral-field spectroscopic
observations of the inner $28''\times28''$ ($3.1 {\rm kpc} \times 3.1
{\rm kpc}$) of the spiral galaxy NGC~5719. The kinematics of the stars
and ionised gas are very complex.
The observed galaxy spectra are decomposed into the contributions of
three distinct kinematic components characterised by a regular
disc-like rotation: one main and one secondary stellar component and a
ionised-gas component.
The spectral decomposition is done using an implementation  of the
pPXF routine, which allows to measure {\it
  simultaneously} the kinematics {\em and\/} stellar population
properties of the two stellar components.

The rotation of the main stellar component is receding towards East,
like that of the stellar body of the galaxy as observed by
\citet{Vergani+07} out to $\sim40''$. We measure a maximum rotation
velocity of $\sim150$ \kms\ at about $10''$  (1.1 kpc) from the centre.
The secondary stellar and ionised-gas components are counter-rotating
with respect to the main stellar component, with a maximum rotation of
$\sim200$ \kms\ at about $10''$ from the centre.
We are able to resolve the two stellar components down to the
innermost $\sim2''$ (0.2 kpc). At smaller radii, they have a too small
velocity separation to be resolved. The median values of light
fraction contributed by the two components in the spatial bins where
they are resolved are $F_{\rm main} = 56\%$ and $F_{\rm secondary} =
44\%$ (with $20\%$ standard deviation), respectively.

The ionised gas is detected all over the observed field of view. 
  It is characterised by a strong \hb\ emission, which is concentrated
  in a twin-peaked morphology indicating an edge-on ring with a
  semi-major axis of $\sim7''$ (0.8 kpc), and an outer asymmetric
  $m=1$ and fragmented spiral arc extending southwards with a
  semi-major axis of $\sim13''$ (1.5 kpc; see contours in
  Figs. \ref{fig:2dkin} and \ref{fig:2dssp}). The two-stream fluid
  instability present in systems like NGC~3593 \citep{Garcia+00} can
  favor this morphology. An hint of the inner ring is visible also in
  the \hi\ position-velocity diagram measured along the major axis of
  NGC~5719 \citep{Vergani+07}.
The secondary stellar component is associated to these features of the
gas distribution, i.e. the counter-rotating stars are detected in a
region enclosed by the ionised-gas structures. A similar phenomenon is
observed in NGC~3593, where a concentrated gaseous ring is associated
to the counter-rotating stellar disc \citep{Corsini+98b, Garcia+00}.

The two counter-rotating stellar components are characterised by
different chemical properties as it results from the measured line
strength of the Lick indices (Fig. \ref{fig:indices}).  On average,
the main stellar component has lower \hb\ and higher [MgFe]$'$, \mgb,
and \fe\ with respect to the secondary component.  This translates
immediately into different properties of their stellar populations
(Fig. \ref{fig:2dssp}).
The main stellar component has ages ranging from 2 to 13.5
Gyr (median $\rm age = 4$ Gyr with 4 Gyr standard deviation). It has
nearly-solar metallicity (median $\rm [Z/H] = 0.08$ dex with 0.20 dex
standard deviation). It displays super-solar enhancement (median $\rm
[\alpha/Fe] = 0.10$ dex with 0.10 dex standard deviation).
The counter-rotating stellar population component is younger, with
ages ranging from 0.7 Gyr to 2.0 Gyr (median age = 1.3 Gyr with 0.6
Gyr standard deviation).  Its metallicity shows a radial gradient: it
changes from sub-solar ($\rm [Z/H] \sim -1.0$ dex) in the outskirts,
to solar ($\rm [Z/H] = 0.0$ dex) and super-solar ($\rm [Z/H] \sim 0.3$
dex) in the centre. The youngest ages and highest metallicities are
found in correspondence of the regions where the \hb\ emission is more
intense. The $\alpha$-enhancement is super-solar (median $\rm
[\alpha/Fe] = 0.14$ dex with 0.05 dex standard deviation).

These findings extend the results by \citet{Neff+05} based on GALEX
observations. They reported the presence of a young stellar component
in the disc of NGC~5719 by analysing UV and optical
images. Unfortunately, the poor GALEX angular resolution prevented
them to associate it to the \hb\ ring as we successfully do
(Fig. \ref{fig:2dssp}). Our estimate is in agreement with their lower
limit to the actual age of such a young population ($0.3\pm0.1$ Gyr).

With our new observations and spectral decomposition technique, we
prove that the mean age of the counter-rotating disc, which is
associated to the neutral and ionised gas disc, is indeed younger than
the main stellar disc. This result shows that counter-rotating disc
has been recently assembled. The median overabundance of the
counter-rotating component (0.14 dex) indicates a star formation
history with a time-scale of 2 Gyr \citep{Thomas+05}.
More details about the assembly process could be derived by a
  further analysis of the east-west asymmetries measured in the maps
  of the stellar-population properties.
The scenario proposed by \citet{Vergani+07} that NGC~5719 hosts a
counter-rotating stellar disc originated from the gas accreted during
the ongoing merging with its companion NGC~5713, is finally confirmed.

\label{lastpage}

\bibliography{coccato11} 
\end{document}